\documentclass[aps,prl,twocolumn,superscriptaddress,longbibliography]{revtex4-2}
\usepackage{amsmath}
\usepackage{graphicx}
\usepackage{hyperref}
\usepackage{color}
\usepackage{amsfonts}
\usepackage{amssymb}
\usepackage{braket}
\usepackage{times}
\usepackage{natbib}
\usepackage{lineno}
\usepackage{siunitx}

\begin{document}

\title{\huge \textbf{Compact linearly uncoupled resonators for efficient spontaneous parametric downconversion via angular phase matching}}

\author{Alessia Stefano}
\email{Contact author: alessia.stefano01@universitadipavia.it} 
\affiliation{Dipartimento di Fisica, Università di Pavia, via A. Bassi 6, 27100 Pavia, Italy
}

\author{Matteo Piccolini}%
\affiliation{Dipartimento di Fisica, Università di Pavia, via A. Bassi 6, 27100 Pavia, Italy
}%

\author{Marco Liscidini}%
\affiliation{Dipartimento di Fisica, Università di Pavia, via A. Bassi 6, 27100 Pavia, Italy
}%

\date{\today}

\begin{abstract}
We report an integrated platform for efficient second-order nonlinear interactions based on linearly uncoupled resonators and angular phase matching. The proposed architecture confines phase control to a limited section of the device, maximizing field enhancement and effective nonlinear interaction length while simultaneously reducing the overall footprint. As an example we show the results for an AlGaAs-on-insulator structure demonstrating a photon-pair generation rate of $3.16~\mathrm{GHz/mW}$ in the continuous-wave regime and $5.89~\mathrm{MHz}$ under pulsed pumping. The generated biphoton state exhibits a Schmidt number $K=1.02$, indicating nearly uncorrelated photon pairs. The compact and reconfigurable nature of this approach, together with its independence from material-specific poling techniques, makes it applicable to a broad class of integrated $\chi^{(2)}$ nonlinear platforms.

\end{abstract}

\maketitle

Integrated sources of entangled photon pairs are fundamental building blocks for a wide range of quantum technologies, including secure quantum communication~\cite{gisin2007quantum}, photonic quantum computing~\cite{ wang2020integrated}, and quantum-enhanced sensing and metrology~\cite{giovannetti2011advances}. These sources are typically based on spontaneous nonlinear processes, such as spontaneous parametric down-conversion (SPDC) or spontaneous four-wave mixing (SFWM), which enable the generation of nonclassical light with high pair generation rates and high purity \cite{yang2008spontaneous, helt2012does}. Among these approaches, SPDC stands out for its broad wavelength tunability, high conversion efficiency, and the availability of mature $\chi^{(2)}$ materials and fabrication platforms \cite{zhu2021integrated}. The continuous development of integrated SPDC sources has led to compact, stable, and scalable devices compatible with existing photonic integration technologies~\cite{guo2017parametric,thiel2024wafer}. In this work, we focus on SPDC-based systems implemented in resonant integrated architectures, where optical confinement and resonance enhancement play a key role in achieving efficient photon-pair generation.

\begin{figure}[t]
    \centering
\includegraphics[width=\linewidth]{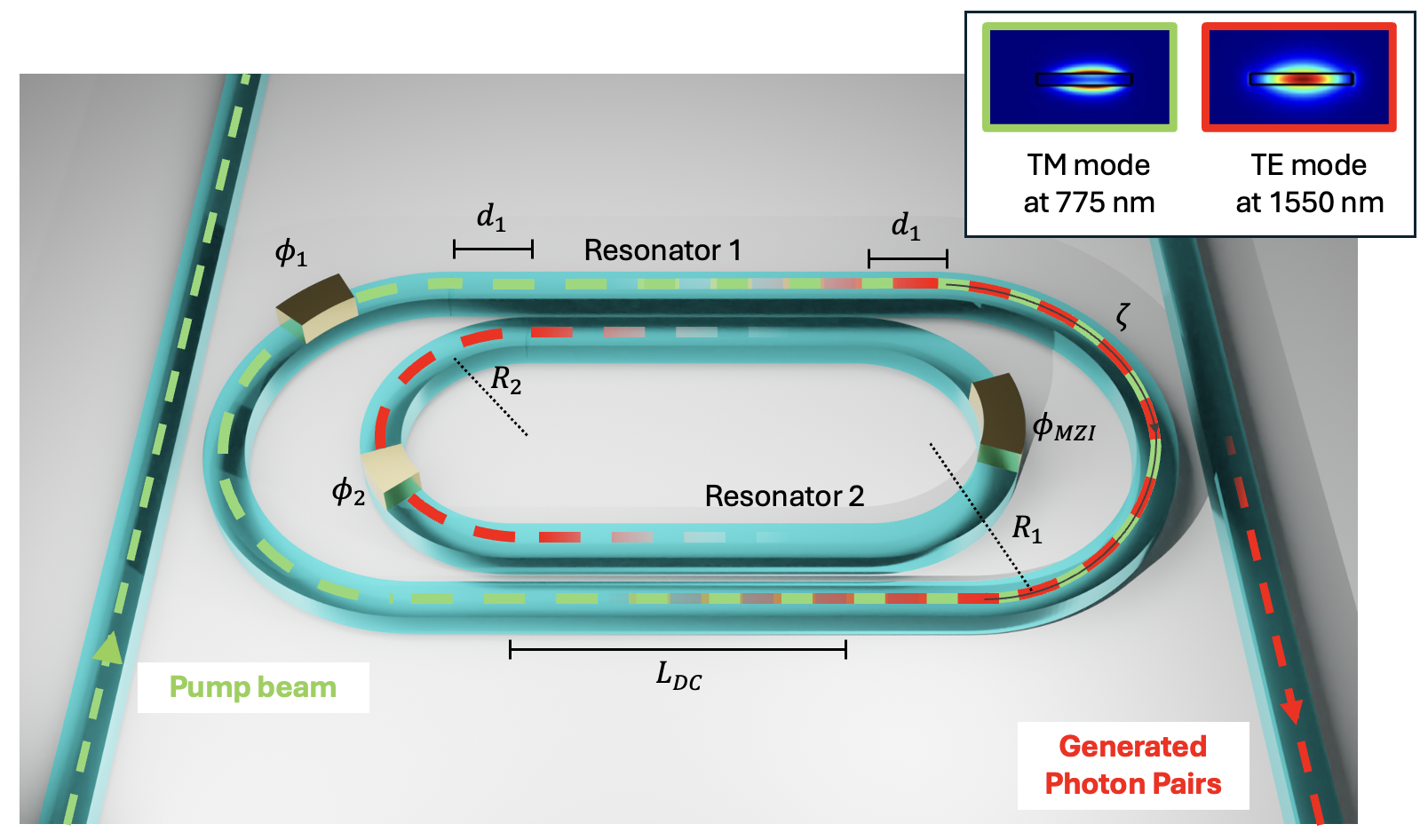}
    \caption{Sketch of the proposed structure. In the inset we show the field distributions for the TM mode at 775 nm and the TE mode at 1550 nm.}
    \label{fig:1}
\end{figure}
In conventional microring resonators, efficient SPDC relies on a doubly/triply resonant condition, where pump, signal, and idler are simultaneously resonant with cavity modes. This is challenging because the modes are nonlinearly coupled over a broad spectral range between the fundamental frequency and its second harmonic. In devices with large quality factors (Qs), in which strong field enhancement is associated with narrow linewidths, resonance alignment becomes highly sensitive to small variations in the ring radius or waveguide geometry. As a result, even when the doubly (or triply)  resonant condition is satisfied at the design stage, unavoidable fabrication imperfections typically shift the resonances from their target positions, requiring additional tuning mechanisms that ultimately limit scalability and yield \cite{lin2016phase, liu2023chi}. Strategies based on careful radius engineering, such as those demonstrated in AlGaAs microrings for SPDC \cite{fontaine2025photon}, can partially mitigate these effects by operating near geometrical configurations with reduced dispersion sensitivity. Nevertheless, in single-ring resonators the doubly resonant condition remains intrinsically fragile.

A similar challenge (although not as severe) had already been identified in the context of SFWM, in which multiple resonance conditions motivated the introduction of linearly uncoupled resonator architectures \cite{menotti2019nonlinear}. The key idea underlying this approach is exploiting resonances associated with different microring resonators, while still allowing for a localized nonlinear interaction in a well-defined spatial region. In this way, the concepts of resonance enhancement and phase matching are effectively disentangled, increasing the structure flexibility. 

Building on these ideas, linearly uncoupled resonator systems have also been extended to SPDC. In these implementations, the nonlinear interaction can be engineered either to take place directly within a directional coupler (DC) connecting the two resonators\cite{zatti2022spontaneous}, or within a periodically poled arm of a Mach–Zehnder (MZ) interferometer\cite{stefano2024broadband}. The former approach already enables independent spectral tuning of the interacting modes, but the periodic poling introduced in the latter configuration along with the MZ-based "uncoupler" provides robustness against fabrication imperfections and allows the doubly resonant condition to be maintained over hundreds of nanometers. This separation of functionalities represents a powerful paradigm, as it allows the recovery of doubly (triply) resonant conditions essentially independently of dispersion and fabrication-induced variations. However, such an approach relies on periodic poling, which limits its applicability across integrated nonlinear platforms. In addition, the designed MZ interferometer requires a structure with a larger footprint and weaker spatial confinement.

In this work, we propose an alternative strategy for implementing linearly uncoupled with phase matching achieved by exploiting the symmetry properties of the second-order nonlinear tensor. Our strategy exploits angular phase matching through the dependence of the effective nonlinearity on the propagation direction of the interacting fields \cite{yang2007enhanced, poveda2023custom}. This mechanism is naturally suited to curved geometries but is not directly compatible with previously studied linearly uncoupled resonator schemes, in which the nonlinear interaction typically occurs along a single propagation direction. To address this, we introduce an alternative structure that combines the advantages of linearly uncoupled resonators with angular phase matching, enabling efficient SPDC without relying on periodic poling. In addition, this is realized in a compact system whose footprint is approximately a factor of two smaller than linearly uncoupled resonators reported to date.

\begin{figure}[t]
    \includegraphics[width=\linewidth]{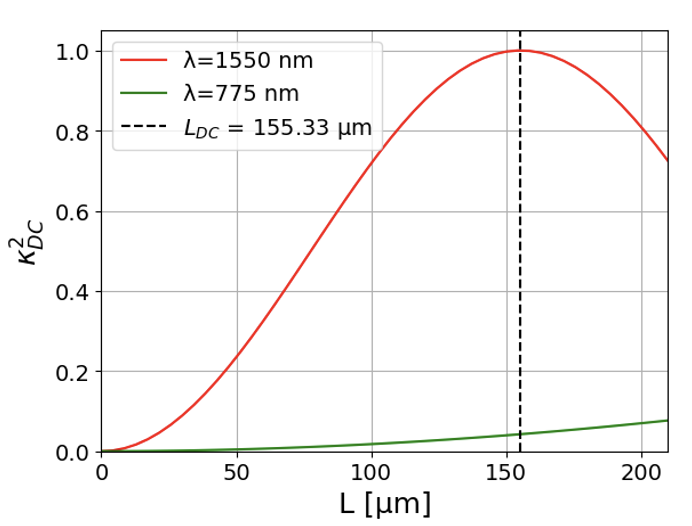}
    \caption{Coupling coefficients of the DC, for the fundamental harmonic (FH, red) and second harmonic (SH, green) as a function of the coupling length.}\label{fig:2}
\end{figure}

The structure considered in this work is schematically shown in Fig.~\ref{fig:1} and consists of two concentric integrated racetracks in AlGaAs embedded in a SiO$_2$ cladding, designed to operate with a pump at the second harmonic, \SI{775}{nm}, and to generate signal and idler photons around the fundamental wavelength, \SI{1550}{nm}. Resonator 1 has a radius $R_1 = \SI{30}{\micro\meter}$ and Resonator 2 has a slightly smaller radius $R_2 = \SI{28.6}{\micro\meter}$. The waveguides feature a rectangular cross section with a height of \SI{105}{nm} and a width of \SI{1100}{nm}, which, together with the choice of $R_1$, is designed to satisfy angular phase matching between the TM0 mode of the pump at \SI{775}{nm} and the TE0 modes of the generated photons at \SI{1550}{nm}. All the modes have been characterized by means of full-vectorial simulations performed using Lumerical \cite{Lum_Mode}. Both racetracks are equipped with independent phase shifters that allow fine control of the supported resonances. Importantly, the two racetracks do not act as independent resonators for all wavelengths; instead, the system is engineered so that the optical modes relevant to the pump and to the generated photons exhibit markedly different spatial distributions. 

At \SI{775}{nm}, the structure supports a mode that is strongly confined to the outer racetrack only, providing field enhancement exclusively in Resonator 1 (see Fig. \ref{fig:1}). In contrast, at \SI{1550}{nm}, the DC gives rise to supermodes whose spatial distribution is for one half of the racetrack in the outer bend, while in the complementary half it is mainly localized in the inner one, as illustrated in Fig.~\ref{fig:1}. This particular modal arrangement plays a central role in enabling the nonlinear interaction. The dominant spatial overlap between the pump and the signal-idler supermodes occurs in the curved section of Resonator 1, where angular phase matching is satisfied. A weaker overlap is also present in the directional-coupler region, but it does not significantly contribute to the nonlinear interaction, as phase matching is not fulfilled there. As a result, the nonlinear interaction is confined to a well-defined portion of the structure, while the remaining sections of the racetracks are  used to independently control the resonances of the interacting fields. 

The excitation and collection scheme further exploits this modal selectivity. The pump at \SI{775}{nm} is injected through a dedicated bus waveguide on the left side of the device, which couples exclusively to the pump resonance. The generated signal and idler photons at \SI{1550}{nm} are instead extracted through a bus waveguide on the right side, which is optimized to efficiently couple the fundamental supermodes. 

The DC plays a central role in controlling the spatial distribution of the FH supermodes. The separation of the racetrack waveguides is \SI{300}{nm} over a length of \SI{155}{\micro\meter} to enable complete coupling at \SI{1550}{nm} while suppressing power transfer at \SI{775}{nm}. In Fig.\ref{fig:2} we show the coupling power coefficients at both the FH and the SH wavelengths as a function of the coupler length. The simulations confirms that  by choosing the length of \SI{155}{\micro\meter} one can achieve complete power transfer at the fundamental wavelength, while suppressing the coupling at \SI{775}{nm} below $0.05$ over a bandwidth of approximately \SI{80}{nm} (see Supplementary Material). Conversely, the coupling coefficient at \SI{1550}{nm} exceeds $0.9$ over a spectral window of about \SI{7}{nm} centered at the target signal and idler wavelengths, confirming the selective operation of the coupler. This design ensures that the pump remains confined to Resonator 1, whereas the generated signal and idler photons are efficiently redistributed between the two paths, realizing the targeted modal structure of the device. 
Finally, to ensure robustness against fabrication defects that may lead to residual field leakage in this section, the structure includes an additional phase shifter, denoted as $\phi_{\mathrm{MZI}}$. By tuning $\phi_{\mathrm{MZI}}$, destructive interference can be enforced between the fields circulating in the right arms of the two racetracks, suppressing unwanted propagation and maximizing the energy density in the regions relevant for nonlinear interaction.

\begin{figure}[t]
    \centering
    \includegraphics[width=\linewidth]{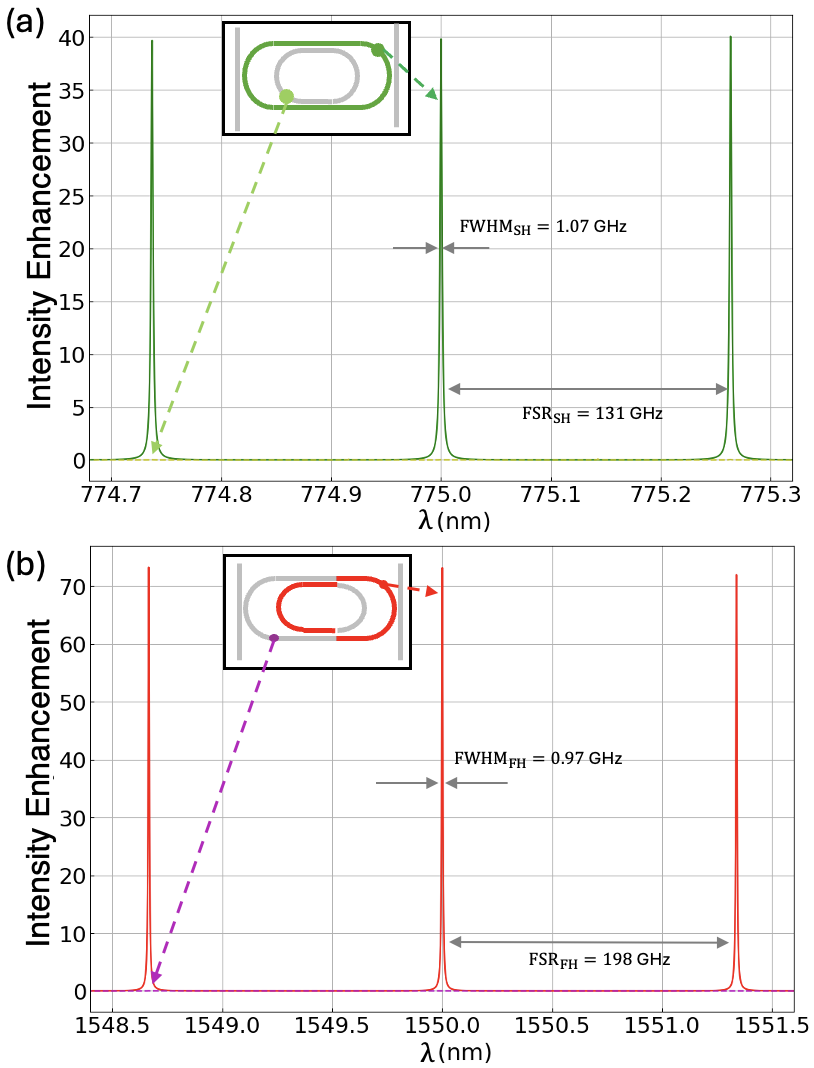}
    \caption{Calculated field enhancement for (a) the SH and (b) the FH. Insets show the device sketch and the regions where the optical fields are evaluated. Dark green (a) and red (b) solid lines indicate the field enhancement in the outer right arm, while light green (a) and purple (b) lines correspond to the inner right arm for FH and to the outer right arm for SH. The vanishing field in these sections confirms the expected behavior of the DC and the effective confinement of the SH field in Resonator 1.}
    \label{fig:3}
\end{figure}

In Fig.~\ref{fig:3} we show the simulated field distributions inside the structure under operating conditions optimized for SPDC, namely critical coupling for pump and slight over-coupling for signal and idler fields, which maximizes the overall generation efficiency. The figure reports the intensity enhancement of the optical fields in different sections of the racetrack for excitation from the corresponding bus waveguides. In particular, in Fig.~\ref{fig:3}(a) we show the SH field intensity when the pump is injected from the left bus waveguide, while in Fig.~\ref{fig:3}(b) we report the FH field intensity for excitation from the right bus waveguide. Solid lines correspond to the intensity enhancement in the outer right arm of the structure, whereas the dotted lines indicate the intensity enhancement in the complementary inner left arms. These results confirm the intended behavior of the DC. In these calculations we consider realistic propagation losses of $\alpha_{\mathrm{FH}} = \SI{1}{dB/cm}$ and $\alpha_{\mathrm{SH}} = \SI{2}{dB/cm}$ \cite{thiel2024wafer}. The total optical path lengths are \SI{501}{\micro\meter} for the FH modes and \SI{511}{\micro\meter} for the SH modes. Under these conditions, the simulated resonance spectra yield intensity enhancement factors of approximately $80$ at \SI{1550}{nm} and $40$ at \SI{775}{nm}. The corresponding free spectral ranges in the two spectral regions are $\mathrm{FSR}_{\mathrm{FH}} = \SI{198}{GHz}$ and $\mathrm{FSR}_{\mathrm{SH}} = \SI{131}{GHz}$, with resonance full widths at half maximum of $\mathrm{FWHM}_{\mathrm{FH}} = \SI{0.97}{GHz}$ and $\mathrm{FWHM}_{\mathrm{SH}} = \SI{1.07}{GHz}$. These values correspond to finesse  of $\mathcal{F}_{\mathrm{FH}} = 204$ and $\mathcal{F}_{\mathrm{SH}} = 123$, and to loaded quality factors of $Q_{\mathrm{L,FH}} = 3.6 \times 10^{5}$ and $Q_{\mathrm{L,SH}} = 1.9 \times 10^{5}$.


Under the assumption of a low generation probability, the two-photon state generated via SPDC in our system can be expressed as~\cite{yang2008spontaneous}:
\begin{equation}
    |\mathrm{II}\rangle = \dfrac{1}{\sqrt{2}} \int d\omega_1\, d\omega_2\, \phi(\omega_1,\omega_2)\, 
    a^\dagger_{\omega_1} a^\dagger_{\omega_2} |\mathrm{vac}\rangle,
\end{equation}
where $|\mathrm{vac}\rangle$ is the vacuum state, and $a^\dagger_{\omega_i}$ is the creation operator for a photon at frequency $\omega_i$. The biphoton wavefunction is given by \cite{yang2008spontaneous}:
\begin{align}
\phi(\omega_1,\omega_2) =&\, i\dfrac{\alpha \bar{\chi}_2}{\beta}
\sqrt{\dfrac{\hbar \omega_S \omega_I \omega_P}{8\pi \epsilon_0 n_S n_I n_P c^3 \mathcal{A}}} \nonumber \\
&\times \phi_P(\omega_1+\omega_2) J(\omega_1,\omega_2,\omega_1+\omega_2),
\end{align}
where $c$ is the speed of light, $\epsilon_0$ the vacuum permittivity, $|\alpha|^2$ the average number of pump photons per pulse, and $|\beta|^2$ the number of generated photon pairs per pulse (with $|\beta|^2 \ll 1$ by assumption). The parameter $\bar{\chi}_2$ denotes the effective second-order nonlinear coefficient of AlGaAs, here assumed to be $100~\text{pm/V}$, while $\mathcal{A}$ is the effective nonlinear interaction area, estimated to be about $ \SI{1}{\micro m^2}$~\cite{yang2008spontaneous}. The effective refractive indices at the pump, signal, and idler frequencies are denoted as $n_P$, $n_S$, and $n_I$, respectively. The pump spectral profile is $\phi_P(\omega)$, and $J(\omega_1,\omega_2,\omega_1+\omega_2)$ denotes the nonlinear field overlap integral~\cite{liscidini2012asymptotic}, defined as:
\begin{align}
J(\omega_1,\omega_2,\omega_1+\omega_2)=&f(\omega_1) f(\omega_2) f(\omega_1+\omega_2)I(\omega_1,\omega_2,\omega_1+\omega_2)
\end{align}
where $f(\omega_i)$ are the field amplitudes in the lower arm of the MZI at frequency $\omega_i$, and $\Delta k=k(\omega_p)-k(\omega_1)-k(\omega_2)\pm2/R_1 $ represents the momentum mismatch along the propagation coordinate $\zeta$. Finally, the function I describes the nonlinear overlap integral along the propagation direction and is
\begin{align}
  I&=\int_0^{\pi R_1} d\zeta e^{i\Delta k\zeta}\cos(\zeta/R_1)\sin(\zeta/R_1) \\
    &=\dfrac{e^{i\pi\Delta k R_1/2}}{4i}\pi R_1\,\mathrm{sinc}[\pi\Delta k R_1/2],\nonumber
\end{align}
where we considered only the term corresponding to the right bend of Racetrack 1, where angular phase matching is satisfied through the proper choice of $R_1$.

We first assess the efficiency of the proposed architecture in terms of photon-pair generation rate. We begin by considering a continuous-wave (CW) pump resonant with the SH mode at frequency $\omega_P = \omega_{\mathrm{SH}}$, where energy conservation enforces $\omega_2 = \omega_P - \omega_1$. Under these conditions, the nonlinear overlap integral depends on a single frequency variable and can be written as $J(\omega_1, \omega_P - \omega_1, \omega_P)$. Following the theoretical framework developed in Ref.~\cite{stefano2024broadband}, we obtain a generation rate per unit pump power of $3.16~\mathrm{GHz/mW}$ per resonance. We then extend the analysis to the experimentally relevant case of a pulsed pump. Specifically, we consider a Gaussian pump pulse with a temporal width of \SI{50}{ps} and pulse energy of $\SI{1.22}{pJ}$, corresponding to a peak power of \SI{10}{mW}. In this regime, the photon-pair generation rate is given by the product of the laser repetition rate and the generation probability per pulse, $|\beta|^2$. Assuming a repetition rate of \SI{10}{MHz}, we obtain an overall generation rate of \SI{5.89}{MHz}, confirming that the proposed architecture enables efficient SPDC in both CW and pulsed excitation.

Our structure can be benchmarked against the implementations reported in Refs.\cite{zatti2022spontaneous} and \cite{stefano2024broadband}. In Ref.\cite{zatti2022spontaneous}, the nonlinear interaction occurs in a DC, while in \cite{stefano2024broadband} it takes place in the arm of an MZI using periodic poling. To compare the performance of these platforms, we introduce an effective nonlinear interaction volume (see Supplementary Material), which quantifies the overall field enhancement provided by each structure in the SPDC process. For the configurations of \cite{zatti2022spontaneous} and \cite{stefano2024broadband}, the effective volumes are $3.8\times 10^4\,\mu\text{m}^3$ and $3.3\times 10^4\,\mu\text{m}^3$, respectively. The structure proposed here has an effective volume of $2.4\times 10^4\,\mu\text{m}^3$, corresponding to a 57\% increase in efficiency with respect to SPDC in the DC and a 36\% improvement over the periodically poled MZI.
Compared to other recently reported architectures, including Ref.\cite{stefano2024broadband}, our structure achieves an approximately twofold reduction in footprint area, while maintaining independent tuning of the interacting fields and fully decoupled coupling sections for the pump and generated photons.

\begin{figure}
    \centering
    \includegraphics[width=\linewidth]{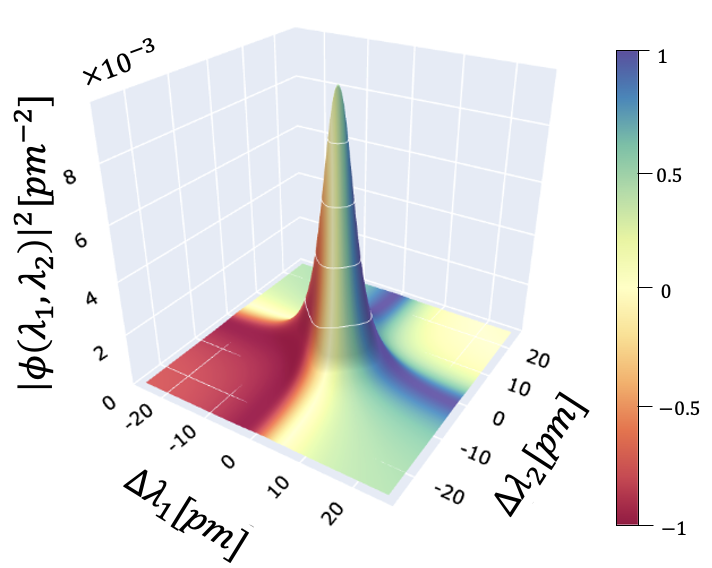}
    \caption{Plot of $
    |\phi(\lambda_1,\lambda_2) |^2[pm^{-2}] $ for the SPDC process in the proposed resonant structure. The color encodes the cosine of the phase. The calculation assumes a Gaussian pump pulse with a temporal length of \SI{49}{ps}, a pulse energy of $\SI{1.22}{pJ}$, and repetition rate of \SI{10}{MHz}. The resulting spectral decomposition yields a Schmidt number $K = 1.02$, indicating nearly uncorrelated photon pairs.}
    \label{fig:4}
\end{figure}


We now turn to the analysis of the spectral correlations of the generated photon pairs. The resulting biphoton wavefunction, computed under the same operating conditions used for the generation-rate analysis in the pulsed regime, is shown in Fig.~\ref{fig:4}. The corresponding Schmidt number is $K = 1.02$, indicating that the generated photon pairs are nearly spectrally uncorrelated. This result is obtained without introducing additional constraints or optimization steps specifically aimed at suppressing correlations~\cite{vernon2017truly}, and while operating in a regime that ensures efficient photon-pair generation. The origin of this behavior can be understood by considering the interplay between the resonance linewidths and the nonlinear interaction mechanism. In SPDC, the process depends linearly on the pump field, and the pump frequency is twice that of the generated photons. As a consequence, spectral separability can be achieved even when the pump quality factor is comparable to those of the fundamental modes, without the need to deliberately engineer strongly mismatched linewidths. This contrasts with SFWM-based approaches, where suppressing spectral correlations typically requires imposing significancy different quality-factor values for the interacting resonances~\cite{vernon2017truly}.


In conclusion, we have presented a reconfigurable integrated platform designed to enhance SPDC by maximizing both field enhancement and effective interaction length. Unlike previously proposed architectures, where a significant portion of the resonator length is dedicated to independent field tuning, our design exploits a compact configuration that maintains tunability while increasing the nonlinear generation region. Numerical simulations confirm efficient phase matching and strong SH field buildup within the interaction section.
We analyzed the SPDC process under both continuous-wave and pulsed pumping, obtaining a generation rate of $3.16~\mathrm{GHz/mW}$ in the CW regime and \SI{5.89}{MHz} for \SI{10}{MHz} repetition rate pulsed excitation. The computed biphoton wavefunction exhibits a Schmidt number $K = 1.02$, demonstrating nearly separable photon pairs suitable for heralded single-photon generation.
While this work focuses on AlGaAs, the angular phase-matching approach is general and can be extended to other nonlinear platforms. In particular, other III–V semiconductors, such as Indium Gallium Phosphide (InGaP) and Gallium Phosphide (GaP) \cite{thiel2024wafer}, as well as non-centrosymmetric materials like Cadmium Sulfide (CdS) \cite{poveda2023custom}, represent promising candidates.
These results validate the potential of our architecture for compact and reconfigurable sources of photon pairs in integrated photonics. Future work will focus on the experimental implementation of the device and its extension to other nonlinear processes such as sum- and difference-frequency generation and on-chip quantum frequency conversion.


\begin{acknowledgments}
\textbf{Acknowledgments}
M.P., A.S acknowledges project PNRR MUR project PE0000023-NQSTI. M.L. acknowledges project PRIN 2022 20222JNHL3-OASIS “Onchip scalable squeezing sources”. 
\end{acknowledgments}

\bibliography{sample}
\end{document}


\title{\huge \textbf{Supplementary Information for ``Compact linearly uncoupled resonators for efficient spontaneous parametric downconversion via angular phase matching"} }%

\author{Alessia Stefano}\email{alessia.stefano01@universitadipavia.it}
\affiliation{Dipartimento di Fisica, Università di Pavia, via A. Bassi 6, 27100 Pavia, Italy}

\author{Matteo Piccolini}
\affiliation{Dipartimento di Fisica, Università di Pavia, via A. Bassi 6, 27100 Pavia, Italy}

\author{Marco Liscidini}
\affiliation{Dipartimento di Fisica, Università di Pavia, via A. Bassi 6, 27100 Pavia, Italy}

\maketitle

\section{Directional Coupler Characterization}

The Directional Coupler (DC) is engineered to efficiently transfer power at the fundamental wavelength while suppressing coupling at the second harmonic. The device geometry is characterized by a coupling gap of \SI{300}{nm} and is optimized using Lumerical eigenmode simulations. For each wavelength and polarization, we define the modal splitting parameter $\Omega_c(\lambda) = \pi\,\Delta N(\lambda)/\lambda,$
where $\Delta N(\lambda) = n_{\mathrm{eff},+}(\lambda) - n_{\mathrm{eff},-}(\lambda).$ This parameter quantifies the supermode separation and determines the spatial oscillation of power exchange in the DC. From coupled-mode theory, the power-coupling coefficient at the FH is
\begin{equation}
    \kappa^2_{\mathrm{DC,FH}}(\lambda, L) = \sin^2\!\left[\Omega_{c,\mathrm{FH}}(\lambda)\,L\right].
\end{equation}
Figure 2 in the main paper reports $\kappa^2_{\mathrm{DC,FH}}$ and $\kappa^2_{\mathrm{DC,SH}}$, plotted in red and green, respectively, as functions of the length of the DC. By enforcing full power transfer ($\kappa^2_{\mathrm{DC,FH}} = 1$) we find the optimal coupler length to be $L_{\mathrm{DC}} = \SI{155}{\micro\meter}$.

\begin{figure}[h!]
    \centering
    \includegraphics[width=0.7\linewidth]{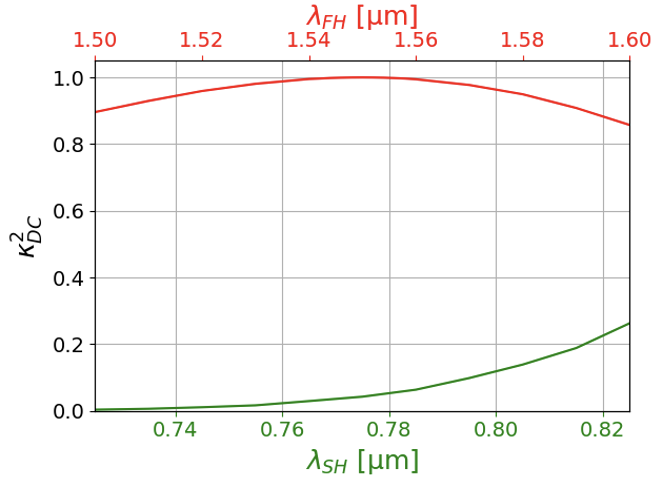}
    \caption{Operational bandwidth of the DCs. The red plot refers to the coupling coefficient $\kappa^2_{\mathrm{DC,FH}}$ of the generated photons at \SI{1550}{nm}, while the green plot refers to $\kappa^2_{\mathrm{DC,SH}}$, for the pump field.}
    \label{fig:1}
\end{figure}

Then, we analyze the bandwidth of the DCs. Numerical simulations were performed using Lumerical \cite{Lum_Mode} to compute the frequency-dependent coupling coefficient $\Omega_c(\lambda)$. Figure \ref{fig:1} shows the resulting DC coupling coefficient for a fixed coupler length $L_{\mathrm{DC}} = \SI{155}{\micro\meter}$ as a function of wavelength.

The upper horizontal axis corresponds to the quantity $\kappa^2_{\mathrm{DC,FH}}$ evaluated in the spectral region around the fundamental harmonic, shown by the red curve. 
The lower horizontal axis refers instead to $\kappa^2_{\mathrm{DC,SH}}$ evaluated around the pump wavelength, shown by the green curve.

From the plot, we observe that $\kappa^2_{\mathrm{DC,SH}} < 0.05$ over a bandwidth of approximately \SI{80}{nm} centered at the pump wavelength $\lambda_P = \SI{775}{nm}$. 
At the same time, the coupling remains strong at the fundamental harmonic, with $\kappa^2_{\mathrm{DC,FH}} > 0.9$ over a bandwidth of approximately \SI{7}{nm} around the target wavelengths $\lambda_S = \lambda_I = \SI{1550}{nm}$. 
These results confirm that the directional coupler operates in the desired regime, enabling efficient interaction for the generated photons while suppressing coupling at the pump frequency.

\begin{table}[h!]
\centering
\caption{\bf Structural efficiency of different SPDC structures based on linearly uncoupled resonators}
\renewcommand{\arraystretch}{1.3}
\begin{tabular}{c c c c}
\hline
Sketch of the structure & Overlap integral & Finesse factor & Effective volume \\
\hline
\includegraphics[width=1.8cm]{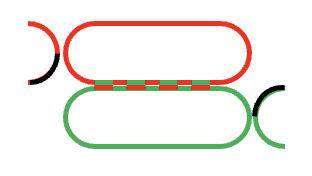} &
$J \simeq \dfrac{L_{\mathrm{DC}}}{4}$ &
$\mathcal{F} \propto \dfrac{1}{L_\mathrm{FH}^2 L_\mathrm{SH}^2}$ &
$V_\mathrm{eff}=3.8 \times 10^4\,\mu\mathrm{m}^3$ \\

\includegraphics[width=1.8cm]{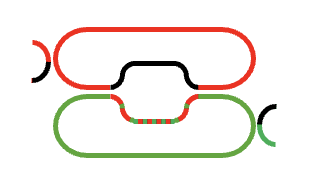} &
$J \simeq L_{\mathrm{MZI}}$ &
$\mathcal{F} \propto \dfrac{1}{L_\mathrm{FH}^2 L_\mathrm{SH}^2}$ &
$V_\mathrm{eff}=3.3 \times 10^4\,\mu\mathrm{m}^3$ \\

\includegraphics[width=1.8cm]{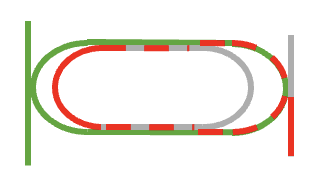} &
$J \simeq L_{\mathrm{int}}$ &
$\mathcal{F} \propto \dfrac{1}{L_\mathrm{FH}^2 L_\mathrm{SH}^2}$ &
$V_\mathrm{eff}=2.4 \times 10^4\,\mu\mathrm{m}^3$ \\
\vspace{0.1cm}\\
\hline
\end{tabular}
\label{tab:shape-functions}
\end{table}

\section{Effective volume as a figure of merit for efficiency}

The probability of generating a photon pair per pump pulse is given by

\begin{equation}
    |\beta|^2 =\, |\alpha|^2 \bar{\chi}_2^2
\dfrac{\hbar \omega_S \omega_I \omega_P}{8\pi \epsilon_0 n_S n_I n_P c^3 \mathcal{A}_{\mathrm{eff}}} |\phi_P(\omega_1+\omega_2) J(\omega_1,\omega_2,\omega_1+\omega_2)|^2,
\end{equation}

where $\alpha$ denotes the number of pump photons. We define the efficiency $\eta$ of the process as $|\beta|^2/|\alpha|^2$. Considering that $J(\omega_{1},\omega_{2},\omega_{1}+\omega_{2})\propto L_{\mathrm{int}}$, where $L_{\mathrm{int}}$ represents the effective interaction length over which the nonlinear process occurs, and by retaining only the contributions arising from the structure geometry, the efficiency can be expressed as

\begin{equation}
    \eta\;\propto\;\frac{L_{\mathrm{int}}^{2}\,\mathcal{F}_{\mathrm{SH}}\,\mathcal{F}_{\mathrm{FH}}^{2}}{\mathcal{A}_{\mathrm{eff}}},
\end{equation}

Here, $\mathcal{F}_{\mathrm{SH}}$ denotes the finesse of the pump resonator, while $\mathcal{F}_{\mathrm{FH}}$ is the finesse associated with the resonator supporting the generated photons. By explicitly writing the finesse terms, the efficiency scaling becomes

\begin{equation}
\eta
\propto
\frac{L_{\mathrm{int}}^{2}}{A_{\mathrm{eff}}}
\left(
\frac{v_{g}^{\mathrm{SH}} Q_{\mathrm{SH}}}{L_{\mathrm{SH}}\,\omega_{\mathrm{SH}}}
\right)
\left(
\frac{v_{g}^{\mathrm{FH}} Q_{\mathrm{FH}}}{L_{\mathrm{FH}}\,\omega_{\mathrm{FH}}}
\right)^{2}.
\end{equation}

We now focus on the terms containing the interaction length, the resonator lengths, and the effective area. Taken together, these quantities have the dimensions of the inverse of a volume. This observation motivates the definition of an effective volume,

\begin{equation}
\frac{1}{V_{\mathrm{eff}}}=\frac{L_{\mathrm{int}}^{2}}{A_{\mathrm{eff}}\,L_{\mathrm{SH}}\,L_{\mathrm{FH}}^{2}}.
\end{equation}

It is important to note that $V_{\mathrm{eff}}$ does not correspond to a physical volume in a geometric sense. Rather, it provides a compact figure of merit useful to compare the efficiency of different structures, accounting not only for spatial confinement through the effective area, but also for the field enhancement induced by the resonant structure.

In general, this formulation shows that the efficiency of the SPDC process is inversely proportional to the effective volume.

In Table 1 we report a comparison between different integrated platforms where SPDC has been investigated. In Ref.~\cite{zatti2022spontaneous}, the nonlinear interaction occurs within the directional coupler, resulting in an effective volume of $3.8\times 10^4 \mu m^3$ . In Ref.~\cite{stefano2024broadband}, the interaction takes place in one arm of a Mach--Zehnder interferometer, leading to an effective volume of $3.3\times 10^4 \mu m^3$ . In the structure proposed in this work, the interaction extends over the entire resonant path, yielding a reduced effective volume of $2.4\times 10^4 \mu m^3$ .

This result benchmarks our design, corresponding to a $57\%$ enhancement in efficiency with respect to the structure reported in Ref.~\cite{zatti2022spontaneous} and a $37\%$ improvement compared to Ref.~\cite{stefano2024broadband}.

\bibliography{sample}
